# Excluded Volume Effect in Unzipping DNA with a Force

Pui-Man Lam[*] , J.C.S. Levy
Laboratoire de Physique Theorique de la Matiere Condensee
Universite Paris 7-Denis Diderot
2 Place Jussieu, 75251, Paris, France

and

Hanchen Huang
Department of Mechanical, Aerospace & Nuclear Engineering,
Rensselaer Polytechnic Institute, Troy, NY 12180

**Abstract:** A double stranded DNA molecule when pulled with a force acting on one end of the molecule can become either partially or completely unzipped depending on the magnitude of the force F. For a random DNA sequence, the number M of unzipped base pairs goes as $M \sim (F-F_c)^{-2}$ and diverges at the critical force $F_c$ with an exponent $\chi=2$. We find that when excluded volume effect is taken into account for the unzipped part of the DNA, the exponent $\chi=2$ is not changed but the critical force $F_c$ is changed. The force versus temperature phase diagram depends on only two parameters in the model, the persistence length and the denaturation temperature. Furthermore a scaling form of the phase diagram can be found. This scaling form is parameter independent and depends only on the spatial dimension. It applies to all DNA molecules and should provide a useful framework for comparison with experiments.

*On leave from Physics Department, Southern University, Baton Rouge, LA 70813
Email: pmlam@grant.phys.subr.edu

## I. INTRODUCTION

In the last decade advances in experimental techniques in atomic force microscopes [1,2], optical tweezers [3,4] and glass microneedles [5,6] have allowed manipulation of single biological molecules, revealing many of their new and unexpected behaviors [7-9]. Of particular interest is the area of DNA molecules where micromanipulation technique involving handles attached to the two ends of the molecule were developed [10] to study its response to external torques [11,12] and its mechanical unzipping in the absence of enzymes [13-15]. The latter topic has only recently been subjected to theoretical investigations[16-21], even though theoretical studies of thermal denaturation of DNA have a much longer history [22-25].

Recently Bhattacharjee [26,27], Lubensky and Nelson [28,29], Bhattacharjee and Marenduzzo [30], and Mukamel and Shakhnovich [31] have extended a minimal model [22] of DNA as two ideal polymer chains, by introducing a pulling force applied to one of the two extremities. By mapping into a non-Hermitian quantum mechanics problem, it was shown in references [26] and [28] that the DNA can be unzipped only if the applied force exceeds a critical value $F_c$. The number of unzipped base pairs M goes as $M \sim (F-F_c)^{-\chi}$ and diverges at the critical force $F_c$, with an exponent $\chi$ which has the value 1or 2 depending on whether the Gaussian polymers are assumed to be homopolymers or heteropolymers respectively. In both cases, the critical force $F_c$ is given by $F_c = 3f_0 \, k_B T/b^2$, where $b^2=2a^2$, with a the Kuhn length of the single-stranded DNA, $k_B$ the Boltzman constant, T the temperature and $f_0$ the average free energy per base pair with no pulling force. Surprisingly the force versus temperature phase diagram revealed the presence [19-21, 29, 31] of a novel reentrant unzipping transition at low temperature.

The result of references [26] and [28] for the physically more interesting case of heteropolymers has recently been rederived using a much simpler model by Chen [32]. Furthermore the model of Chen can be generalized to incorporate the freely joint chain (FJC) model [33] which takes into account the finite extensibility of the chain, while in references [26] and [28] the chain can be stretched without bound under a strong pulling force. One of us has shown that the FJC model can be used to incorporate the effect of excluded volume of the chain [34]. Here we will study the effect of excluded volume on the unzipping of DNA using the freely joint chain model. We find that the exponent $\chi=2$ is not changed by excluded volume effect while the critical force $F_c$ is changed. The force versus temperature phase diagram depends on only two parameters in the model, the persistence length P and the denaturation or melting temperature $T_D$. Furthermore, a universal scaling form for the phase diagram can be found. This scaling form is parameter independent and depends only on the spatial dimension. It applies to all DNA molecules and should provide a convenient framework for comparison with experiments. Although the typical experiment is in the strong force limit in which excluded volume effect is not important, there is evidence that in the general case [35], self-avoidance can be relevant. In section II we will review the model of Chen and show how it can be generalized to incorporate the FJC model. In section III we will discuss how to incorporate excluded volume effect in the FJC model and its effect on the unzipping of DNA. In section IV we discuss the special case of two dimensions in which the effect of excluded volume is even stronger. Section V is the conclusion.

**II.    A Mean Field Theory**

In the theory of Chen [32], the DNA molecule with N base pairs, being pulled apart by a force F at one of its ends, has (M/2) base pairs unzipped, the average distance between the two separated ends being z. It is assumed that the averaged free energy per base pair is $-f_0$, with $f_0 > 0$ when the DNA is below the melting temperature with no pulling and the interaction energy per monomer of the unzipped part can either be $+\varepsilon$ (repulsive) or -ε (attractive). The reduced free energy Γ of the entire chain can be written as (with β=1/k$_B$T)

$$\beta\Gamma = -\left(N - \frac{M}{2}\right)\beta f_0 + \frac{3z^2}{2Mb^2} - \beta F z - \beta\varepsilon\sqrt{M} \qquad (1)$$

The first term describes the fact that there are (N-M/2) base pairs that are still in the bound state on average. The second term comes from the Gaussian statistics of the polymer, which consists of M monomers since each base pair gives rise to two monomers when unzipped. The third term is simply a potential energy reduction of the terminal pairs when an external pulling field F is applied. The fourth term comes from the randomness of the heteropolymer sequence [36].

Minimizing the free energy with respect to z leads to

$$z = b^2\beta FM/3 \qquad (2)$$

Substituting this value of z in (1) we have

$$\Gamma = -Nf_0 + tM - \varepsilon\sqrt{M} \qquad (3)$$

where we have defined a reduced force parameter

$$t = f_0/2 - \beta F^2 b^2/6 \qquad (4)$$

From (3) one can see that for $t \leq 0$, there is no minimum of the free energy as the number M of unzipped monomers tends to infinity and the double stranded DNA becomes entirely unzipped. However, for $t > 0$, there is an equilibrium state given by minimizing the free energy with respect to M. The average number of unzipped pairs is given by

$$M = \varepsilon^2/(4t^2) \qquad (5)$$

and the free energy is

$$\Gamma = -Nf_0 - \varepsilon^2/(4t) \qquad (6)$$

The unzipping transition takes place at t = 0, which is equivalent to saying that there is a critical unzipping force $F_c$ given by

$$\frac{F_c b}{k_B T} = \sqrt{3f_0/(k_B T)} \qquad (7)$$

The results (5)-(7) are consistent with those of Lubensky and Nelson [28] that M diverges with an exponent $\chi = 2$ as F approaches $F_c$. From (2) one can see that the polymer can be stretched without bound with a strong force. This unphysical feature can be corrected by using the FJC model. In this model the free energy is given by [33]

$$\Gamma_{FJC} = -Mk_BT\ln\{\sinh(\beta Fb)/(\beta Fb)\} \qquad (8)$$

Replacing the second and third terms of (1) by (8) we have

$$\Gamma = -(N-M/2)f_0 - Mk_BT\ln\{\sinh(\beta Fb)/(\beta Fb)\} - \varepsilon\sqrt{M} \qquad (9)$$

With this form of the free energy, equations (5) and (6) remain unchanged in form. The only change is the definition of t, which now reads

$$t = f_0/2 - k_BT \ln\{\sinh(\beta Fb)/(\beta Fb)\} \qquad (10)$$

The finite extensibility of the chain is now taken into account. The critical force $F_c$ is now given by the solution $t = 0$ in (10). The exponent $\chi = 2$ remains unchanged. However the statistics of the chain is still Gaussian and no excluded volume effect has been taken into account.

### III. Excluded Volume Effect

In order to take into account the excluded volume effect, we will make use of the model of Orlandini et al [19]. In this model the free energy is given by

$$\Gamma = -\left(N - \frac{M}{2}\right)f_0 - k_BT \log[P_M(z)\mu_u^M \mu_z^{N-\frac{M}{2}}] - Fz \qquad (11)$$

where $P_M(z)$ is the probability that an M-step chain has an end-to-end distance z and $\mu_u, \mu_z$ are the effective coordination of the unzipped and zipped parts of the chain, respectively. In section II we had neglected the contribution of the effective coordination to the entropy. Therefore the model in section II does not possess a finite melting temperature $T_D$ which is the temperature at which the critical force vanishes. In this section we take this effect into account. This will give rise to a finite melting temperature $T_D$.

It is well known [37-39] that

$$P_M(z) \sim \exp\left\{-\frac{d}{2}\left[\frac{z}{<R>}\right]^\delta\right\} \qquad z >> <R>$$

where $<R>$ the average end-to-end distance, $\delta = 1/(1-\nu)$, with $\nu$ the usual correlation length critical exponent [37], and d is the spatial dimension. This distribution reduces to the Gaussian distribution when $\delta = 2$ and $\nu = 1/2$. Defining the persistence length P as

$$P = <R>^2/(2M^{2\nu}a) \qquad (12)$$

where a is the monomer length, this quantity is a constant for both the Gaussian chain and the chain with excluded volume since in both cases $<R> \sim M^\nu$. This scaling form for the end-to-end distribution is correct only for small stretching force F. For large force the length z should be proportional to M and we will have to go to the freely-joint-chain model to correct for this. Substituting this definition of P into $P_M(z)$ we have

$$P_M(z) \sim \exp\left\{-\frac{d}{2}\left[\frac{z}{\sqrt{2PM^{2\nu}a}}\right]^\delta\right\} \qquad z >> M^\nu a \qquad (13)$$

Substituting (13) into (11), the free energy has the form

$$\Gamma = k_B T \frac{d}{2}\left[\frac{z}{\sqrt{2PM^{2\nu}a}}\right]^\delta - \frac{Mk_B T}{2}\log\left[\frac{\mu_u^2}{\mu_z}\right] - Fz + \frac{M}{2}f_0 - Nf_0 + \text{constant} \qquad (14)$$

Minimizing the free energy $\Gamma$ with respect to z yields

$$z = \left[\frac{2F\sqrt{2PM^{2\nu}a}}{d\delta k_B T}\right]^{1/(\delta-1)} \qquad (15)$$

Substituting this value of z back into (14) gives the free energy in the form

$$\Gamma = M\left[\frac{f_0}{2}\left(1-\frac{T}{T_D}\right) - (2Pa)^{\frac{1}{2\nu}}(\delta-1)\left(\frac{2F^\delta}{d\delta^\delta k_B T}\right)^{\frac{1}{\delta-1}}\right] + \text{constant} \qquad (16)$$

where the melting temperture $T_D$ is given by

$$T_D = \frac{f_0}{\log\left(\frac{\mu_u^2}{\mu_z}\right)} \qquad (17)$$

The critical force $F_c$ for unzipping is obtained when the term proportional to M in (16) vanishes. This can be written in the form

$$\frac{2F_c\sqrt{2Pa}}{f_0(1-(T/T_D))} = \frac{\delta}{(\delta-1)^{(\delta-1)/\delta}}\left(\frac{dk_B T}{f_0(1-(T/T_D))}\right)^{\frac{1}{\delta}} \qquad (18)$$

From (15) we can see that the extension z increases without bound with F. In order to correct this we can make use of the freely-joint chain model. Now from (9) the free energy of FJC model under force F, taking into account the full entropy due to the effective coordinations $\mu_u$, $\mu_z$ has the form

$$\Gamma = -(N-M/2)f_0 (1-(T/T_D)) - Mk_BT\log\{\sinh(\beta Fb)/(\beta Fb)\} - \varepsilon\sqrt{M} \quad (19)$$

Expanding the second term of (19) containing the logarithm for small F, one finds

$$\Gamma_{FJC} = -M\frac{F^2 b^2}{6k_B T} \quad (20)$$

Equating this to the second term inside the bracket of (16) with d=3, one finds an expression for the Kuhn length b

$$b = \left[\frac{2Pa(\delta-1)}{\delta^{\frac{\delta}{\delta-1}}}\right]^{1/2} 2^{\frac{1}{(\delta-1)}} \left(\frac{3k_B T}{F\sqrt{2Pa}}\right)^{\frac{\delta-2}{2(\delta-1)}} \quad (21)$$

We know that for large F the Kuhn length b reduces to that of the standard freely joint chain result $b=\sqrt{2Pa}$, since in that case the effects of excluded volume must disappear when the chain is almost fully stretched. Our equation (21) does not have this property but it does give a cutoff force $F_{cut}$ above which the excluded volume effect is unimportant, given by

$$F_{cut} = \frac{3k_B T}{\sqrt{2Pa}} \quad (22).$$

We can now define an effective Kuhn length $b_{eff}$ as

$$b_{eff} = \left[\frac{2Pa(\delta-1)}{\delta^{\frac{\delta}{\delta-1}}}\right]^{1/2} 2^{\frac{1}{(\delta-1)}} \left(\frac{F_{cut}}{F}\right)^{\frac{\delta-2}{2(\delta-1)}} [1-\exp(-mF_{cut}/F)] + \sqrt{2Pa}\exp(-mF_{cut}/F)$$

$$(23)$$

where m is a parameter of order 1. This effective Kuhn length reduces smoothly to b given by (21) at small F and to the freely joint chain result $\sqrt{2Pa}$ at large F. The parameter m can be varied until the resulting critical force for unzipping to be determined below do not change appreciably.

We can approximately take into account the effect of excluded volume by substituting this expression of the effective Kuhn length $b_{eff}$ into the freely-joint chain free energy (19). The critical force $F_c$ is again given by the condition for the vanishing of the term linear in M in (19):

$$\exp\left[\frac{f_0(1-(T/T_D))}{2k_BT}\right] = \frac{\sinh(\omega)}{\omega} \qquad (24)$$

where

$$\omega = Fb_{eff}/k_BT \qquad (25)$$

with $b_{eff}$ given by (23).

Equations (23) to (25) can be solved for the scaled quantity $F_c(2Pa)^{1/2}/[f_0(1-(T/T_D))]$ as a function of the scaled quantity $k_BT/[f_0(1-(T/T_D))]$. The parameter in (23) is determined to be 2 or 3. The result is shown in Figure 1, together with that of equation (18), with d=3, $\delta$=2.5 and $\nu$=0.6 [37]. The results of the standard freely-joint chain, obtained with $\delta$=2 and $\nu$=1/2 and the Gaussian chain result obtained from (7) are also shown for comparison. One can see that the result of the freely-joint chain model with excluded volume effect approaches that of (18) at large T. Similarly, the standard freely joint chain result approaches that of the Gaussian chain in that limit. We emphasize here that in Figure 1 we have shown a universal scaling form for the phase diagram. This scaling form is parameter-independent and depends only on the spatial dimension. The different models shown in Fig. 1 represent different approximations to a universal scaling function which should apply to all DNA molecules characterized by a persistence length P, average free energy per base pair $f_0$ and a melting temperature $T_D$. We would also like to point out that our results are correct both for very small and very large force F. For intermediate values of F our results are only approximate. Therefore it is not likely that our scaling function is correct for all F. However, in spite of this, it is likely that the scaling plot, i.e. $F_c(2Pa)^{1/2}/[f_0(1-(T/T_D))]$ versus $k_BT/[f_0(1-(T/T_D))]$ would still produce a universal scaling function for all experimental data with different DNAs, since everything is expressed in reduced units and all quantities, the persistence length P, monomer length a, base-pair binding energy $f_0$ and melting temperature $T_D$ are all measurable quantities. The persistence length P, however should be calculated according to equation (12) in the case where self-avoidance is important. This should provide a convenient framework to compare with experiments.

The behaviors of the critical force $F_c$ in the limits $T \rightarrow T_D$ can be easily obtained. In the limit $T \rightarrow T_D$ we can expand the right hand side of (22) for small $\omega$ and find the exact behavior given in (18), with d=3, in agreement Orlandini et al [19]. We find, for $T \rightarrow T_D$ the critical force vanishes as

$$\frac{2F_c\sqrt{2Pa}}{f_0} = \frac{\delta}{(\delta-1)^{\frac{\delta-1}{\delta}}}\left(\frac{3k_BT}{f_0}\right)^{\frac{1}{\delta}}\left(1-\frac{T}{T_D}\right)^{\nu} \qquad (26)$$

In Figure 1 we have shown a scaling form of the phase diagram, independent of the parameters P and $T_D$ of the model. Experimentally, it is known that the denaturation temperature of DNA is about 350K and the average free energy per base pair is given by $f_0 \approx 2.5 k_B T$ [40,41]. Using this information it is possible to obtain the critical force as a function of the temperature. In Figure 2 we show the the quantity $F_c (2Pa)^{1/2}/(k_B T)$ versus the temperature T, for the four different cases of FJC with excluded volume, the model of Orlandini et al, the standard FJC and the Gaussian chain.

**IV Special case of two dimensions**

The case of two dimensions is of interest because of the stronger effect of excluded volume in lower dimensions. Such low dimensional geometry can in principal be achieved experimental by confining the DNA molecule between two large slabs.

For two dimensions, the free energy is given by

$$\Gamma = -(N-M/2)f_0 (1-(T/T_D)) - M k_B T \log\{I_0(\beta Fb)\} - \varepsilon \sqrt{M} \qquad (27)$$

where $I_0$ is the modified Bessel function of order zero, with small F behavior $I_0(\beta Fb) \approx 1 + (\beta Fb/2)^2$. Comparing the small F behavior of (27) with (16) for d=2 we obtain

$$b = 2 \left[ \frac{2Pa(\delta-1)}{\delta^{\frac{\delta}{\delta-1}}} \right]^{1/2} \left( \frac{k_B T}{F\sqrt{2Pa}} \right)^{\frac{\delta-2}{2(\delta-1)}} \qquad (28)$$

from which we can see that the cutoff force for the freely joint chain behavior is given by

$$F_{cut} = \frac{k_B T}{\sqrt{2Pa}} \qquad (29)$$

We can therefore define an effective Kuhn length $b_{eff}$ as

$$b_{eff} = 2 \left[ \frac{2Pa(\delta-1)}{\delta^{\frac{\delta}{\delta-1}}} \right]^{1/2} \left( \frac{F_{cut}}{F} \right)^{\frac{\delta-2}{2(\delta-1)}} [1-\exp(-mF_{cut}/F)] + \sqrt{2Pa} \exp(-mF_{cut}/F) \qquad (30)$$

where m is a parameter of order 1 which can be determined as in the three dimensional case. We can take into account the effect of excluded volume by substituting (30) for the value of the effective Kuhn length. The critical force $F_c$ is given by the condition

$$\exp\left[ \frac{f_0(1-(T/T_D))}{2k_B T} \right] = I_0(\omega) \qquad (31)$$

with $$\omega = Fb_{eff}/k_BT \qquad (32)$$

Equations (30) to (32) can be solved for the scaled quantity $F_c(2P)^{1/2}/[f_0(1-(T/T_D))]$ as a function of the scaled quantity $k_BT/[f_0(1-(T/T_D))]$. The parameter m in (30) is determined to be 2 or 3. The result is shown in Figure 3, together with that of equation (18), with d=2, $\delta$=4 and $\nu$=3/4, known exactly in two dimensions[37]. The results of the standard freely-joint chain, obtained with $\delta$=2 and $\nu$=1/2 and the Gaussian chain result obtained from (7) are also shown for comparison. One can see that the result of the freely-joint chain model with excluded volume effect approaches that of (18) at large T. Similarly, the standard joint chain result approaches that of the Gaussian chain in that limit. Again we can interpret the results of Fig. 3 as different approximations to the universal scaling function which is now different than in the three dimensional case.

Using $f_0$ =2.5 $k_BT$ and $T_D$ =350K as in the three dimensional case we have plotted the critical force $F_c(2P)^{1/2}/(k_BT)$ versus the temperature T, for the four different cases of FJC with excluded volume, the model of Orlandini et al, the standard FJC and the Gaussian chain.

## V  Conclusion

We have studied theoretically the effect of excluded volume on the unzipping of DNA by pulling the double strands apart at one end with a force F. We find that the number of unzipped base pairs M goes as $M\sim(F-F_c)^{-\chi}$. We find that the exponent $\chi$=2 is not changed by excluded volume effect. However the critical force $F_c$ is changed. The force versus temperature phase diagram depends on only two parameters in the model, the persistence length and the denaturation temperature. Furthermore, a universal scaling form for the phase diagram can be found. This scaling form is parameter independent and depends only on the dimension. It applies to all DNA molecules characterized by a persistence length P, average free energy per base pair $f_0$ and a melting temperature $T_D$. This should provide a convenient framework for comparison with experiments.

**Aknowledgement.** P.M. Lam would like to thank D. Marenduzzo and A. Maritan for very useful discussions, the Laboratoire Physique Theorie de la Matierie Condensee, Universite Paris 7 and the Department of Mechanical Engineering, The Hong Kong Polytechnic University (PolyU 1/99C) for hospitality.

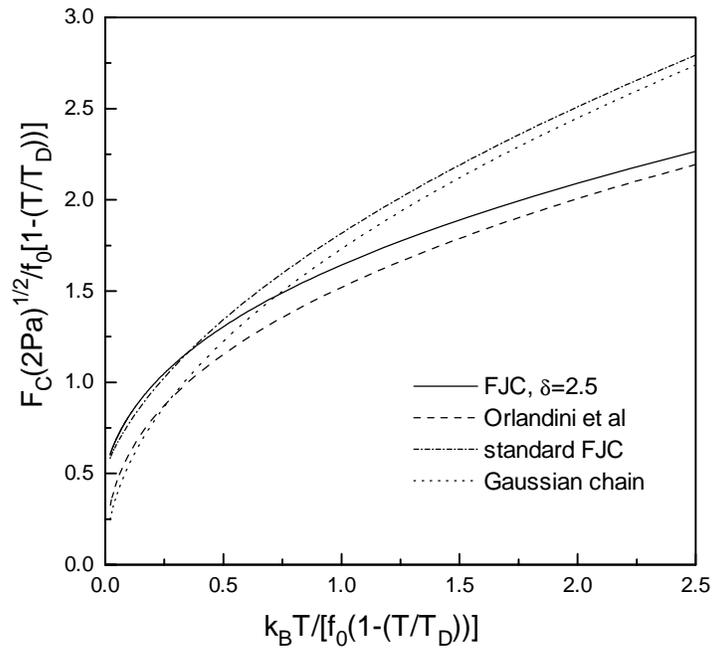

Figure 1.

Figure Captions:

Figure 1: Scaled quantity $F_c(2Pa)^{1/2}/[f_0(1-(T/T_D))]$ versus the scaled quantity $k_B T/f_0 (1-(T/T_D))]$, for various models in three dimensions.

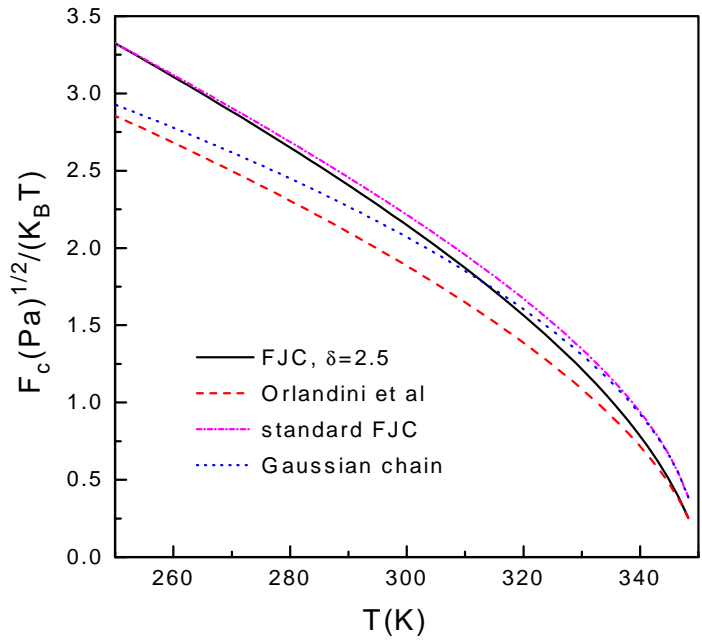

Figure 2. Force versus temperature phase diagram obtained using $f_0 = 2.5\, k_B T$ and $T_D = 350K$, for various models in three dimensions.

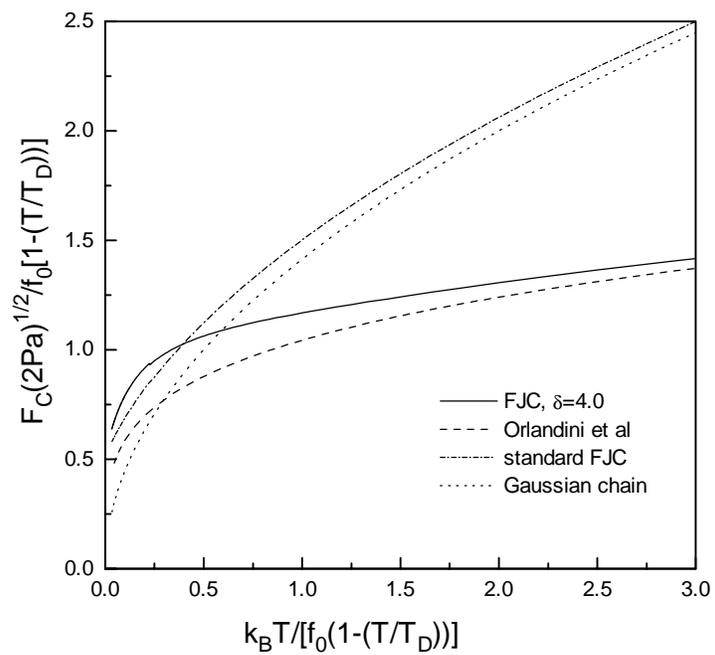

Figure 3. Same as Fig. 1, but for two dimensions.

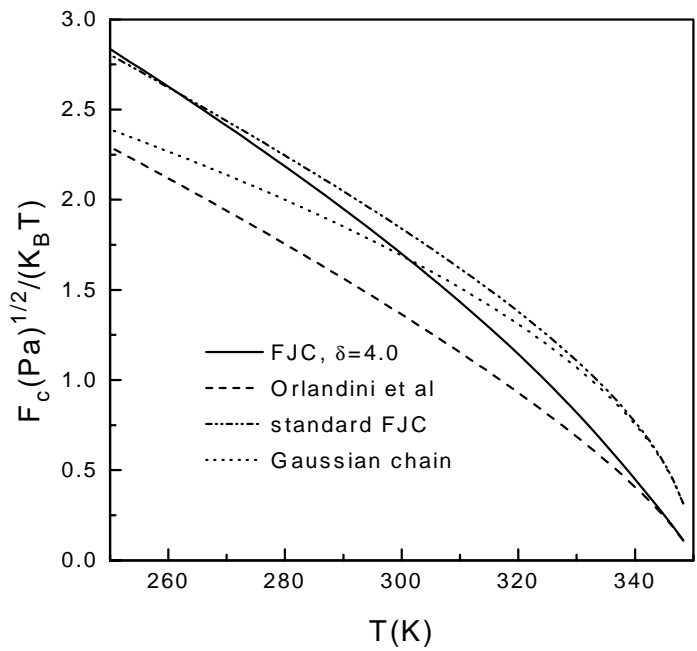

Figure 4. Same as Fig. 2, but for two dimensions.